\documentstyle[prl,aps,amsmath,amsfonts,latexsym]{revtex}

\begin{document}
\title{Are there Thermodynamical Degrees of Freedom of Gravitation?}

\author{H.-H. v. Borzeszkowski 
and T. Chrobok}
\address
{Institut f\"{u}r Theoretische Physik, Technische Universit\"{a}t
Berlin, Hardenbergstr. 36, D-10623 Berlin, Germany}

\maketitle

\begin{abstract}
In discussing fundamentals of general-relativistic irreversible
continuum thermodynamics, this theory is shown to be characterized by
the feature that no thermodynamical degrees of freedom are ascribed
to gravitation. However, accepting that black hole thermodynamics
seems to oppose this harmlessness of gravitation one is called on
consider other approaches. Therefore, in brief some gravitational and
thermodynamical alternatives are reviewed.
\end{abstract}

\section{Introduction}

Due to the equivalence of energy, inertia and gravitation, the
gravitational fields are universally coupled to all physical matter
and systems, respectively. As a consequence, gravitation cannot be
switched off or screened such that one has always to regard as well
the influence of external gravitational fields as of the
gravitational self-field of the system under consideration. Thus, the
laws of thermodynamics and general relativity theory must be applied
simultaneously. And to ensure that this provides a self-consistent
physical description one has to unify or, at least, to harmonize
these laws. The fact that gravitational fields are generally so weak
that one can neglect their action does not free oneself from this
task since, first, one needs its solution for an exact description
and, second, in cosmology and astrophysical objects like neutron
stars the gravitational field is strong.

The standard version of general-relativistic irreversible
thermodynamics results as an adaption of the non-relativistic
continuum theory of irreversible processes near the equilibrium
(non-relativistic TIP) to general relativity theory (GRT); therefore,
let us call it general-relativistic TIP. It is reached by following
that rule which is given by the above-mentioned principle of
equivalence: First one has to look for a special-relativistic version
of the theory under consideration (this provides Lorentz-covariant
basic laws formulated in the Minkowski space-time), afterwards one
has to transit to the general-relativistic formulation with
coordinate-covariant laws formulated in a Riemannian space-time,
whose curvature tensor measures the gravitational field.

This straight road from non-relativistic TIP to general-relativistic
TIP does not seem to encounter any hindrance. Our point, however, is
that the theory established in this manner only leads to a
harmonization of irreversible thermodynamics and GRT. It retains its
physical justification only by presupposing that the gravitational
field has no thermodynamical degrees of freedom. However, this
assumption is questionable for two reasons: (i) The resulting
general-relativistic thermodynamics misses most of that content which
it started from and which legitimizes it as thermodynamics. Its
general-relativistic generalization made TIP a physically void
scheme. This loss concerns the first law of thermodynamics and the
hypothesis of local equilibrium (and, thus, also the second law of
thermodynamics). (ii) The thermodynamics of black holes seems to show
that gravitation has thermodynamical degrees of freedom.

Therefore, general-relativistic TIP (and, as will be shown, certain
of its generalizations) has to be reconsidered in order to win hints
at a theory representing a genuine unification of irreversible
thermodynamics and gravitational theory. As a result of this, all
speaks for a thermodynamics far from the equilibrium or/and a
relativistic theory of gravitation which, in contrast to GRT,
maintains the notion of energy\footnote{Possibly, the unification of
thermodynamics and general relativity can only be reached in the
statistical frame. But, nevertheless, also the fluid approximation of
statistical thermodynamics has to be compatible with GRT.}.

Of course, thermodynamicists and relativists know a variety of
arguments in favor of a modification of their respective theories,
TIP and GRT. Even more, there are conceptions for such modifications
and also elaborated alternative theories. Thus, the purpose of this
paper is to add a further argument stemming from a critical
investigation of general-relativistic TIP.

The paper is organized as follows. To remind of the original physical
content of continuum thermodynamics we intend to show as missing in
its general-relativistic version, in Sec. \ref{thermo}, the
transition from non-relativistic TIP to special-relativistic TIP is
briefly summarized. In Sec. \ref{thermogr}, we turn to
general-relativistic TIP and some of its extensions, discuss the
problems of unification one meets therein, and propose possible
solutions of the problem.
\section{From the non-relativistic field theory of irreversible processes (non-relativistic TIP)
 to its special-relativistic version (special-relativistic TIP)} \label{thermo}
Non-relativistic TIP (cf., e.g. \cite{DEGROOT}) is a field theory
at which one arrives by reformulating classical (quasi-static)
thermodynamics in two steps: First one reinterprets the basic
relations of classical thermodynamics (first and second laws,
Carnot-Clausius relation, Gibbs law), originally formulated in the
state space, as relations in space and time (for arguments, see
\cite{TRUESDELL}). Accordingly, one has to replace thermodynamical
potentials by space-time functions and 1-forms (or differentials)
of quantities defined in the state space by the material (or
substantial, or co-moving) derivative, where
$d/dt=\partial/\partial t+v^{\mu}\partial_{\mu}$ (small Greek
indices run from 1 to 3). This provides the relations
($U=$internal energy, $Q=$heat supply, $W=$work, $S=$entropy,
$\Theta=$temperature):
\begin{eqnarray}
\mbox{{\it{First law of thermodynamics:}}} \qquad
\frac{dU}{dt}=\frac{dQ}{dt}+\frac{dW}{dt}
\end{eqnarray}
\begin{eqnarray}
\mbox{{\it{Second law of thermodynamics:}}} \qquad
\frac{dS}{dt}=\sigma\geq0
\end{eqnarray}
\begin{eqnarray}
\mbox{{\it{Carnot-Clausius theorem ($\overset{o}{}$ denotes the
equilibrium value):}}}\quad
\Theta\frac{d\overset{o}{S}}{dt}=\frac{dQ}{dt}
\end{eqnarray}
\begin{eqnarray}
\mbox{{\it{Gibbs law:}}} \qquad\frac{d
U}{dt}=\Theta\frac{dS}{dt}+\frac{dW}{dt}
\end{eqnarray}
In a second step, these integral relations are rewritten in the
form of local laws by introducing densities and fluxes of the
extensive quantities. Following this procedure one arrives at
non-relativistic TIP which is based on the balances for mechanical
quantities and the first and the second laws of thermodynamics.
Thus, these laws take the general form \cite{MUSCHIK2001}:
\begin{eqnarray}
\frac{\partial u_A}{\partial t}+\partial_{\mu}\phi^{\mu}_A=r_A
\end{eqnarray}
where $u_A(x^{\mu},t)$ are the wanted fields, the
$\phi^{\mu}_A(x^{\mu},t)$ the fluxes, $r_A(x^{\mu},t)$ the supply and
production terms. They are differential equations for the wanted
fields and valid for arbitrary materials. From them one can again
return to their integral formulations. These relations must be
supplemented by constitutive equations which are defined on the state
space and depend on the specific material under consideration.
Inserting them into the balance relation one obtains balances on the
state space.

For this theory stems from classical thermodynamics it maintains
its basic assumption, namely the hypothesis of local equilibrium
saying in particular (see, e.g., \cite{MUSCHIK2001}) that (i) the
state space of non-relativistic TIP is the equilibrium state
space, (ii) the "volume elements" are considered as systems in
local equilibrium ("Schottky systems"), and the values of the
fields do not change in a volume element, but differ only from
volume element to volume element thus describing a non-equilibrium
situation.

This means also that, in non-equilibrium, the time-derivative of
the entropy satisfies the Carnot-Clausius theorem and, thus, the
Gibbs law ($\rho c^2=n(mc^2+\epsilon)$, where $\epsilon$ is the
internal energy and $n$ the particle density; $\mu$ denotes the
chemical potential):
\begin{eqnarray}
\frac{d(ns)}{dt}=\frac{1}{\Theta}\frac{d(\rho
c^2)}{dt}-\frac{\mu}{\Theta}\frac{d n}{dt}
\end{eqnarray}
Now this law holds not just for transitions between equilibrium
states but for arbitrary infinitesimal displacements. Assuming the
Casimir-Onsager relations the Gibbs law together with the balance
equations for mass and internal energy enables one to calculate those
fluxes and forces which determine the entropy production.

Turning to the special-relativistic generalization of TIP
\cite{ECKART,%
LANDAU,%
WEINBERG,%
EHLERS,%
ISRAEL,%
NEUGEBAUER,%
ISRAELNEU,%
ISRAELNEU2}, one has to replace the above-given basic relations by
such ones that are covariant with respect to global Lorentz
transformations. Accordingly, the basic quantities have to be Lorentz
tensor fields (we choose the signature $-2$). The intensive
quantities temperature $\Theta$, pressure $p$, chemical potential
$\mu$ are associated to scalars measured by an observer at rest with
respect to the fluid, the extensive quantities entropy, volume,
particle number are associated to vectors $S^a$, $u^a$, $N^a$, and
energy and momentum to a (symmetric) tensor $T^{ab}$ (small Latin
indices run from $0$ to $3$). These quantities have to satisfy the
relations\footnote{In \cite{NEUGEBAUER}, in contrast to eq.
(\ref{10}), the entropy vector is assumed to be
$S^a=-\beta_aT^{ab}-\alpha N^a$.}:
\begin{eqnarray}
\mbox{{\it{Balance equations:}}}\qquad \nabla_a N^a=0\\
\label{EISRT} \nabla_aT^{ab}=0
\end{eqnarray}
\begin{eqnarray}
\mbox{{\it{Dissipative equation:}}}\qquad \nabla_aS^a=\sigma\geq0
\end{eqnarray}
\begin{eqnarray} \label{10}
\mbox{{\it{Carnot-Clausius condition:}}}\qquad
S^a=\phi^a-\beta_bT^{ab}-\alpha N^a
\end{eqnarray}
\begin{eqnarray} \label{11}
\mbox{{\it{Gibbs law:}}} \qquad dS^a=-\alpha dN^a-\beta_bdT^{ab}
\end{eqnarray}
where $u_a$ is the 4-velocity, $\beta_a=u_a/\Theta$ the inverse
temperature vector, $\phi^a=p\beta^a$ the fugacity vector, and
$\alpha=\mu/\Theta$. It is assumed that the equilibrium
energy-momentum tensor $\overset{o}{T^{ab}}$ is given by that one of
ideal matter and $\alpha$ and $\beta^2=\beta_a \beta^a=c^2/\Theta^2$
are equal to their equilibrium values.

In global thermal equilibrium, the thermodynamical gradients are
constrained by
\begin{eqnarray} \label{KILLING}
\partial_a\alpha=0 \qquad \nabla_a\beta_b+\nabla_b\beta_a=0
\end{eqnarray}

This approach to special-relativistic TIP\footnote{The theories of
Eckart \cite{ECKART} and Landau and Lifshitz \cite{LANDAU} are
typical representatives of this approach.
(For a comparison of them, see \cite{ISRAELNEU3,%
DEGROOTNEU}.)} is characterized by the assumption that
off-equilibrium thermodynamics is described by the same variables and
with the same entropy current as in equilibrium thermodynamics. This
framework can be generalized by assuming a "second-order" scheme
where
\begin{eqnarray} \label{13}
S^a=p(\alpha,\beta)\beta^a-\alpha N^a-\beta_bT^{ab}-Q^a(\delta
N^c,\delta T^{cd}, X_A^{c...})
\end{eqnarray}
where $Q^a$ is of second order in the derivatives $\delta N^c,\delta
T^{cd}$ and generally depends also on other variables $X_A^{c...}$
that vanish in equilibrium (in first-order theory one had $Q^a=0$).
Specific versions of this scheme generally considered in
\cite{ISRAELNEU2} are extended thermodynamics theories
\cite{ISRAELNEU3,%
MULLER1993,%
JOU,%
ISRAELNEU4} and divergence thermodynamic theories
\cite{LIU,%
GEROCH,%
CALZETTA}.\footnote{Since the specific form of $S^a$ is not
relevant for our further arguments we shall not comment its
different versions, either in first-order or second-order
theories.}

Special-relativistic TIP and its generalizations given by (\ref{13})
are also thermodynamical theories near the equilibrium based on the
Carnot-Clausius relation and the Gibbs law. Similar as in
non-relativistic TIP, it is again possible to return to integral
balance relations which, for closed systems, become conservation
laws.

Regarding volume forces this scheme can be extended. For instance, in
the case of electrodynamical forces the above-given basic Minkowski
tensors have to be supplemented by the electromagnetic excitation
tensor $H^{ab}$ ($H^{ab}=-H^{ba}$), the electromagnetic field
strength tensor  $F_{ab}$ ($F_{ab}=-F_{ba}$), and the basic relations
have to be modified by adding the energy-momentum tensor of the
electromagnetic field to $T^{ab}$. Furthermore, the balance equations
must by supplemented by the electromagnetic field equations
(containing the electric current vector $J^a$):
\begin{eqnarray}
\nabla_bH^{ab}=J^a \qquad \nabla_{[a}F_{bc]}=0
\end{eqnarray}
The Carnot-Clausius relation remains unchanged if one assumes that
the electromagnetic field does not contribute to the entropy current
(cf, e.g., \cite{NEUGEBAUER}).
\section{On General-relativistic TIP and some of its generalizations} \label{thermogr}
If gravitational fields are incorporated the equivalence principle
forbids one to do this in the same way as in the electromagnetical
case. Instead, one has to regard gravitational effects by
"lifting" all equations from the Minkowski into the Riemann
space-time. This means to replace in the equations of
special-relativistic TIP the 4-metric $\eta_{ab}$ by the 4-metric
$g_{ab}$ and the partial 4-derivative $\nabla$ by the covariant
derivative $D$.

This procedure is unique because the lifting procedure can be
performed before the constitutive equations have to be taken into
consideration, i.e., on a level where still all relations to be
lifted are differential equations of first order. (Problems occur
for differential equations of higher-order because the covariant
derivatives do not commute so that then one encounters a
derivative-ordering problem.)

In general-relativistic TIP, the basic fields are Riemann tensors
and the basic relations read as follows:
\begin{eqnarray}
\mbox{{\it{Balance equations:}}}\qquad D_a N^a=0\\ \label{EI}
D_aT^{ab}=0
\end{eqnarray}
\begin{eqnarray} \label{18}
\mbox{{\it{Dissipative equation:}}}\qquad D_aS^a=\sigma\geq0
\end{eqnarray}
\begin{eqnarray}
\mbox{{\it{Einstein equations:}}}\qquad G_{ab}=\kappa T_{ab}.
\end{eqnarray}
Furthermore, one requires again the Carnot-Clausius relation
(\ref{10}) and the Gibbs law (\ref{11}).
\subsection{Problems of general-relativistic TIP}
(1) The local balance equations generally cannot be rewritten in an
integral form. As a consequence, the local "conservation" laws (first
of all, the first law of thermodynamics) are no genuine conservation
laws. This is due to the fact that there is no energy-momentum tensor
of gravitation \cite{EINSTEIN1916,%
EINSTEIN1918,%
BERGMANN,%
 MOLLER1966}. Indeed, the Gau\ss{} theorem allowing to transit
from local divergence equations to (integral) conservation laws holds
in the Riemann space only for the divergence of vectors. In the
latter case one has
\begin{eqnarray} \label{31}
D_aA^a=0\Rightarrow\int_{V_4}D_aA^ad^4x=0\\ \label{GAUSS}
\int_{V_4}D_aA^ad^4x=\int_{\partial
V_4^{(1)}}A_adf^a+\int_{\partial V_4^{(2)}}A_adf^a=0
\end{eqnarray}
($V_4$ denotes a 4-dimensional volume, i.e., a 4-cylinder, and
$\partial V_4$ two space-like interfaces of the cylinder) such that
one obtains
\begin{eqnarray} \label{33}
\int_{\partial V_4´}A_adf^a=constant
\end{eqnarray}
saying that the quantity is independent of the interface $\partial
V_4$. But for tensors of higher rank, e.g., for tensors of second
rank one has
\begin{eqnarray}
D_aA^{ab}=0\Rightarrow\int_{V_4}D_aA^{ab}d^4x=0
\end{eqnarray}
which, due to a missing Gau\ss{} theorem, cannot be rewritten in a
form similar to (\ref{GAUSS}). This implies that eq. (\ref{EI})
is no conservation law \cite{EINSTEIN1916,%
EINSTEIN1918,%
BERGMANN,%
 MOLLER1966}. The missing
energy-momentum law is due to the above-mentioned fact that there
does not exist an energy-momentum tensor of gravitational fields.
This becomes obvious when one writes the latter relation (\ref{EI})
equivalently as
\begin{eqnarray}
D_aT^{ab}=\nabla_a(T^{ab}+t^{ab})=0
\end{eqnarray}
Thus, the covariant divergence is transposed into an ordinary
divergence what enables one to apply the Gau\ss{} theorem. But it
does not help because $t^{ab}$ is no tensor; it is the so-called
gravitational energy-momentum complex. Only in equilibrium, where,
due to (\ref{KILLING}), one has a time-like Killing vector field
$\beta_a$ eq. (\ref{EI}) becomes the energy conservation law.

As an implication of the missing law of energy- momentum
conservation, there is no first law in general-relativistic
thermodynamics. In special-relativistic thermodynamics, one obtains
this law by multiplying the law of energy-momentum conservation
(\ref{EISRT}) with the velocity $u_b$ (for details, see
\cite{DEGROOTNEU}):
\begin{eqnarray} \label{ENERGY}
u_b\nabla_aT^{ab}=0
\end{eqnarray}
With the expression of the energy-momentum tensor,
\begin{eqnarray}
T^{ab}=c^{-2}\rho u^au^b+c^{-2}(q^au^b+q^bu^a)+ph^{ab}+\pi^{ab}
\end{eqnarray}
where
\begin{eqnarray}
\rho:&=&c^{-2}u_au_bT^{ab}\qquad \mbox{(energy density)}\\
q^a:&=&h^a{}_cT^{bc}u_b=I^a+h h^{ab}N_b\qquad \mbox{(heat flow)}\\
P^{ab}:&=&h^a{}_cT^{cd}h^b{}_d=p h^{ab}+\pi^{ab} \qquad
\mbox{(pressure tensor)}\\ h^{ab}:&=&\eta^{ab}-c^{-2}u^au^b \qquad
\mbox{(projector orthogonal to $u_a$)}\\ h:&=&\frac{\rho+p}{n} \qquad
\mbox{(specific enthalpy)}
\end{eqnarray}
Eq. (\ref{ENERGY}) provides the first law in the form
($\nabla=N^a\nabla_a$):
\begin{eqnarray} \label{FIRSTSRT}
\nabla e+p\nabla n^{-1}&=&\pi^{ab}\nabla_b u_a-\nabla_a I^a+(h
n)^{-1}I^ah_{ab}\nabla^b p
\end{eqnarray}
This law shows that the change of energy $e$ is produced by two
work and two heat terms.

In general-relativistic thermodynamics, instead of (\ref{EISRT}),
one has (\ref{EI}). Therefore, one obtains instead of Eq.
(\ref{FIRSTSRT}) the relation ($D:=N^aD_a$):
\begin{eqnarray}
D e+p D n^{-1}&=&\pi^{ab}D_b u_a-D_a I^a+(h n)^{-1}I^ah_{ab}D^b p
\end{eqnarray}
Because of the missing physical interpretation of eq. (\ref{EI}) as a
local energy-momentum-conservation, it cannot be interpreted as first
law in the original spirit of thermodynamics, but as one of the
equations of motion.

Now one could oppose that this is nothing but the well-known
problem that in GRT there is no law of energy conservation and
that one has to live with it in all branches of GRT. Why not in
general-relativistic thermodynamics, either? Our point, however,
is that here, because the energy conservation is a basic law of
thermodynamics, this problem is more serious than in other
branches of physics. If, for instance, one considers the influence
of gravitation on a Dirac field then one has also to lift the
Dirac equations in the Riemann space-time. Of course, then the
notion of energy loses its meaning, too. However, in contrast to
general-relativistic TIP and some of the above-mentioned
extensions, by this one does not lose basic laws determining the
dynamics of the physical system.

(2) The loss of the notion of energy leads to problems with the
entropy, too. It is true that the dissipative relation (\ref{18})
contains the divergence of a vector such that the reformulation
described in Eqs. (\ref{31})-(\ref{33}) is possible, but without a
notion of energy, in the framework of TIP, the notion of entropy is
physically meaningless.

(3) The basic assumption of TIP, namely the hypothesis of local
equilibrium (including Carnot-Clausius relation and the Gibbs law),
makes no sense in GRT since there are no Schottky systems.

(4) One meets problems with the interpretation of the metric and
Einstein's equations that can be summarized by the following
questions (for a discussion of the first two questions, see also
\cite{HEEMAYER}). (i) Is the metric an independent variable of the
state space, or is it given by a constitutive equation? (ii) What
about Einstein's equations? They must be solved simultaneously with
the balances. But are they balance equations? (iii) The Einstein
equations imply $D_aT^{ab}=0$. Therefore, one has to ask whether in
GRT room is left for a genuine first law.

Usually general-relativistic TIP and its extensions are justified by
considering such volume cells $d \Sigma_b$ (arbitrarily oriented
elements of 3-areas) that the integral of the $4$-momentum flux
$dp^a=c^{-1}T^{ab}d\Sigma_b$ should vanish to an order higher than
the $4$-volume, when evaluated in locally inertial coordinates (where
$\partial_a g_{bc}=0$) \cite{ISRAELNEU2}. Then all relations reduce
to their special-relativistic versions which can be interpreted in
the original spirit of thermodynamics. In particular, one has
infinitesimal Schottky systems that allow one to define local
equilibrium. That means, the above-discussed general-relativistic
thermodynamics is based on the assumption that the gravitational
field $g_{ab}$ has no thermodynamical degrees of freedom. This
assumption harmonizes but does not unify both theories. However, this
seems to contradict black hole thermodynamics (see, e.g.,
\cite{FROLOV} and the literature therein) showing that the
gravitational field can be ascribed an entropy.
\subsection{Some possible way outs of the dilemma}
(1) One maintains general-relativistic TIP and considers
thermodynamical problems only for special solutions having a
time-like Killing vector or a conformal Killing vector field. In
these cases one can define energy and formulate a corresponding
conservation law. For instance, in Friedman-Robertson-Walker
universes exist time-like conformal Killing vector fields what
allows one to establish cosmological thermodynamics.

(2) One maintains GRT and replaces non-relativistic TIP by a
non-equilibrium thermodynamics requiring neither conservation laws
nor the hypothesis of local equilibrium.

(3) Alternatively to (2), one maintains non-relativistic TIP and
replace GRT by a generalized GRT providing conservation laws
(e.g.,
the teleparallized GRT \cite{EINSTEIN1928,%
MOLLER2}) and unify both fields.

(4) One combines the modifications of TIP and GRT. 

(5) Under relativists it is a popular idea to consider black hole
thermodynamics as guideline to a general-relativistic
thermodynamics (see, e.g., \cite{FROLOV}). This idea is
interesting for it enables one, at least for some special
solutions of Einstein's equations, to geometrize thermodynamical
quantities like temperature and entropy. As mentioned above, in
contrast to general-relativistic TIP, where these quantities are
only ascribed to the system moving in the gravitational field but
not to this field itself, here thermodynamical quantities are
ascribed to the gravitational field, too. However, it is difficult
to see how black hole thermodynamics can be connected with
ordinary thermodynamics. In \cite{NEUGEBAUERNEU} it is attempted
to build a bridge between them by deriving a "universal parameter
thermodynamics for rotating fluids".

(6) Another possibility is to incorporate the entropy as a primary
quantity given by the gravitational field. This idea goes back to
Penrose \cite{PENROSE} who recognized that the Weyl curvature should
vanish at the initial singularity (this is sometimes called {\it Weyl
curvature conjecture}). Following this idea several measures for the
gravitational entropy were formulated (see, e.g., \cite{GRON} and the
references therein). All these measures are build from curvature
invariants such that for vanishing Weyl curvature the gravitational
entropy also vanishes. The advantage of this attempt is that the
gravitational field obtains thermodynamical degrees of freedom, but
the problem of the missing first law still remains open.

(7) Moreover, one may find a formulation of Einsteins equations
which, at least partly, has the structure of divergence equations.
These equations can be included in the scheme of divergence
thermodynamic theories. Physically this would mean to ascribe the
gravitational field thermodynamical degrees of freedom (a paper
concerning this topic is in preparation).



\end{document}